\documentclass[%
reprint, 
superscriptaddress,
amsmath,amssymb,
prx,
floatfix,
]{revtex4-2}

\usepackage{lineno}
\usepackage{hyperref}
\usepackage{csquotes}
\usepackage{circledsteps}
\usepackage{graphicx}
\usepackage{dcolumn}
\usepackage{bm}
\usepackage{ulem}
\renewcommand{\emph}[1]{\textit{#1}}

\usepackage{setspace}
\usepackage[capitalise]{cleveref}
\usepackage{physics}
\usepackage{xspace}

\newcommand{\suppmat}{Supplemental Material\xspace}

\newcommand{\suppfigyoungsmodulus}{Supp.~Fig.~S2\xspace}
\newcommand{\suppfigvelocitydecomposition}{Supp.~Fig.~S3\xspace}

\newcommand{\suppfigspeedheatmap}{Supp.~Fig.~S6\xspace}
\newcommand{\suppfigspeedcorrelation}{Supp.~Fig.~S7\xspace}
\newcommand{\suppfigmsd}{Supp.~Fig.~S8\xspace}
\newcommand{\suppfigmsdslope}{Supp.~Fig.~S9\xspace}
\newcommand{\suppfiginwardsmovement}{Supp.~Fig.~S10\xspace}
\newcommand{\suppfigmixingheatmaploglog}{Supp.~Fig.~11\xspace}
\newcommand{\suppfigaacf}{Supp.~Fig.~S12\xspace}
\newcommand{\suppfigtumblingabp}{Supp.~Fig.~S13\xspace}
\newcommand{\suppfigactivevelocityfit}{Supp.~Fig.~S14\xspace}
\newcommand{\suppfigbetafitparam}{Supp.~Fig.~S15\xspace}
\newcommand{\suppfigvelocityonset}{Supp.~Fig.~S16\xspace}

\newcommand{\suppvidgrowth}{Supp. Video 1\xspace}
\newcommand{\suppvidsmixing}{Supp.\ Videos 2 to 5\xspace}

\newcommand{\removed}[1]{}

\newcommand{\cs}{~}

\renewcommand{\vec}[1]{\boldsymbol{\mathrm #1}}
\newcommand{\tens}[1]{\underline{\boldsymbol{\mathrm #1}}}

\newcounter{myfigpanel}[figure]

\newcommand{\panelletter}[1]{\refstepcounter{myfigpanel}\label{#1}\alph{myfigpanel}}
\newcommand{\panel}[1]{(\protect\panelletter{#1})}
\newcommand{\silentpanel}[1]{{\protect\refstepcounter{myfigpanel}\label{#1}}}
\makeatletter

\newcommand{\@fragilepanels}[3]{
(\panelletter{#1}--
\@for\@pref:={#2}\do{\silentpanel{\@pref}}%
\panelletter{#3})
}%
\newcommand{\panels}{\protect\@fragilepanels}

\newcommand{\mycaption}[1]{\caption{\setcounter{myfigpanel}{0}#1}}

\crefalias{myfigpanel}{figure}

\makeatother

\newcommand{\MPIDS}{\affiliation{Max Planck Institute for Dynamics and Self-Organization, Göttingen, Germany}}
\newcommand{\IDCS}{\affiliation{Institute for the Dynamics of Complex Systems, Göttingen University, Göttingen, Germany}}

\def\figurewidth{\columnwidth}

\begin{document}

\title{Motility-induced mixing transition in exponentially growing multicellular spheroids}%

\author{Torben Sunkel}
\MPIDS
\IDCS
\author{Lukas Hupe}%
\MPIDS
\IDCS
\author{Philip Bittihn}
\email{philip.bittihn@ds.mpg.de}
\MPIDS
\IDCS

\begin{abstract}

Growth drives cellular dynamics in dense aggregates including bacterial colonies, developing tissues, and tumors.
We investigate the underlying physical principles emerging from the interplay of growth, steric repulsion, and motility in a minimal agent-based model of exponentially growing, three-dimensional spheroids.
Our results reveal a motility-induced mixing transition: Without motility, deterministic radial motion from volume expansion dominates, while growth and division cause tangential, cellular-scale diffusion, largely independent of expansion velocity.
Despite this small-scale diffusion, cell lineages remain confined to their local environment.
This confinement persists at weak motility and is overcome only above a threshold, leading to tangential superdiffusivity and global cell mixing with a diverging timescale near the transition, reminiscent of glassy dynamics.
Using a phenomenological model, we identify two effects governing this transition: Steric interactions that suppress motility-induced velocity below a threshold, and the expanding nature of the system which inhibits complete mixing.
Our study highlights the complex interaction of local cell division and motility with global expansion, mediated exclusively by mechanics.
The results provide a baseline for identifying additional biological mechanisms in experiments, for example in tissue spheroids.
The mixing dynamics might also be relevant for competition or tumor progression by interacting with genetic heterogeneity.

\end{abstract}

\maketitle

\section{Introduction}

Biological systems composed of many cells are prime examples of active matter due to their inherent non-equilibrium activity, which is fueled by energy consumption at the microscopic scale. Self-organization in these systems can be driven by a number of different activities which are also present in non-biological systems, including chemical activity\cs\cite{TodaSignaling2018,LiebchenSynthetic2018} and self-propulsion or motility\cs\cite{Golestanian2013, GompperRoadmap2020}.

Growth as an activity is a unique feature of living systems and, at the same time, ubiquitous. Due to its indispensable role in organizing processes as diverse as biofilm formation, embryogenesis or tumor development, proliferating active matter has received increasing attention in recent years\cs\cite{Kuhl14,Hallatschek2023}.
Depending on the properties of the growing system and its constituents, proliferation can lead to distinct emergent phenomena. For example, for anisotropic particles with nematic properties, growth can lead to the formation of locally or globally aligned domains with peculiar mechanical properties\cs\cite{DellArciprete2018, You2018, IsenseeStressAnisotropy2022, Langeslay2023} and to topological defects, which generically appear in active nematics and are important for the morphogenesis of tissues and bacterial colonies\cs\cite{guillamat22,Sarkar2023,Doostmohammadi2016, DellArciprete2018}. Proliferation is also known to drive morphological changes such as the transition to the third dimension\cs\cite{You19, Hartmann2019}, patterning due to stress sensitivity\cs\cite{weady2024selfinhibitinggrowth} or growth-induced entanglement of branching substrates\cs\cite{Day2024}.

Growth often appears together with other processes such as chemical activity, either explicitly or implicitly due to inherent metabolism, whose effects on the environment may in turn regulate growth itself in addition to mechanical interactions\cs\cite{bittihnEngineeredPhenotypicStructure2020,Basan2009homeostatic,Mateus2021growthsignaling,Kuhl14}. Together with cell removal, this feedback can lead to turnover of cells in well-defined homeostatic states, which has been shown to enable competition and mechanical fluidization\cs\cite{Ranft2010,pollack22}. Although unbalanced growth with or without regulation has been considered in specific contexts\cs\cite{Tjhung2020,DellArciprete2018,Doostmohammadi2016}, the effect of unconstrained exponential growth -- as it occurs before the onset of regulation in tissue cultures, bacterial colonies or tumors -- on colony mechanics and its interaction with other activities has received less attention.

Therefore, here, we take a complementary approach and consider pure unbalanced growth in the absence of any regulation or removal. In three dimensions, any growing assembly of particles starting from a small seed will naturally develop into a spheroidal colony. Despite this simplistic morphology, it is important to note that there exists an analogous experimental model system:
Already proposed decades ago\cs\cite{sutherland84, MuellerKlieser1987, holtfreter43, Schwach91}, multicellular spheroids are highly reproducible, approximately spherical colonies of cells that can be produced in bioreactors\cs\cite{azevedo21} and whose complexity ranks between the ones of \textit{in vitro} cellular monolayers and \textit{in vivo} samples\cs\cite{Han2021, Pampaloni2007}. They are becoming a standard model system for drug development\cs\cite{azevedo21, Schueler22, Rossi2022} as well as for biophysical research, with the recent discovery of pressure-driven outbursts in 3D cancer aggregates\cs\cite{Raghuraman2022}, curvature-induced velocity waves in rotating spheroids\cs\cite{Brandsttter2023}, and the drive of cancer cell invasion in the surrounding extracellular matrix (ECM) by adhesive cancer-asscoiated fibroblasts (CAFs) in spheroids\cs\cite{Labernadie2017}, to only name a few examples. Consequently, spheroids have recently also been studied experimentally and theoretically from the perspective of growing active matter, establishing connections to several of the above-mentioned phenomena such as alignment\cs\cite{Ruske2023}, heterogeneous and glass-like dynamics\cs\cite{Sinha2020} and fluidization\cs\cite{Malmi18}.

While relevant biological realizations such as tissue spheroids, tumors or bacterial aggregates do indeed exist, our aim is not to model a specific one of these systems. Instead, by considering idealized physical limits, we hope to establish a clearly defined class of systems to study fundamental processes and to extract unique features compared to non-growing systems without volume expansion. The results could therefore be useful not just for cellular aggregates, but even for the future design of artificial self-organized growing materials.

To this end, we introduce an exclusively mechanical model where a single cell is represented by a dumbbell of spheres which separate over the course of the cell cycle. This idea has been employed before \cite{Drasdo2005, pollack22} and its opportunities have recently been discussed by Hupe and Pollack et al.\cs\cite{Hupe2024}. To establish a baseline, we first consider growth and division as the only drivers of activity, while steric repulsion mediates the intercellular interactions and working near the incompressible limit ensures homogeneous expansion.
We characterize the resulting dynamics using directional mean squared displacements of cells which separate expansion-flow driven motion from the remaining dynamics and reveal diffusive tangential motion. We then endow cells with a motility mechanism that ensures force balance and thus physical consistency by employing reciprocal intercellular forces. This allows us to investigate how the additional activity of cells actively moving through the colony alters the dynamics. We find several features, including superdiffusive tangential motion at a certain time scale and large-scale mixing, which only emerge above a critical motility at which the mixing time scale diverges. We build a simple phenomenological model that describes the mixing dynamics and allows theoretical insights into the interaction of different processes contributing to the mixing transition, and find two responsible mechanisms based on the dense and expanding nature of the spheroidal system. Some aspects of the phenomenology resemble a glass transition, that has been found in biological systems\cs\cite{Malmi18, Bi2016, Angelini2011, Boocock2023} or soft matter in general\cs\cite{Berthier2019, Tjhung2020, Ikeda2012} before. Finally, to elucidate the interaction between the different kinds of activities, we investigate the influence of different growth rates on the mixing dynamics.

\section{Model}

Our model can be understood as a 3D version of the disk cell model presented by Hupe and Pollack et al.\cs\cite{Hupe2024}. We model cells as dumbbells of two spherical \textit{nodes} of unit diameter that continuously grow and divide, as illustrated in \cref{pan:division}.
For each cell, this life cycle is parameterized by its growth progress $g \in [0,1]$, which increases with a constant rate $\gamma=\dd g/\dd t$.
Node centers are connected by a spring of rest length $b^\mathrm{eq}$ along the backbone vector $\mathbf{b}$ (\cref{pan:forceinternal}). 
To make the cell expand, $b^\mathrm{eq}$ increases linearly with $g$: At $g = 0$, both nodes overlap completely; as the cell ages, they are pushed apart until finally detaching and forming two new cells at $g = 1$.

To prevent global synchronization of cell cycles, the growth rates $\gamma$ of these new daughter cells are drawn randomly from a uniform distribution over the interval $[\alpha-\frac{\alpha}{4}, \alpha+\frac{\alpha}{4}]$. 
Here, $\alpha$ is the bulk doubling rate of the colony\cs\footnote{Note that due to statistical effects discussed in detail by Isensee et al.\cs\cite{IsenseeSupplement2022}, the actual average over cell growth rates can be smaller than $\alpha$ depending on the width of the distribution. Here, this statistical correction amounts to a factor of approximately $0.985$}, corresponding to exponential growth with rate $\alpha\,\ln(2)$.
For simplicity, we first set $\alpha$ to $1$, such that 1 time unit corresponds to the average life time of a cell, or 1 generation. However, we will later also vary the growth rate, where this simple relationship will therefore not hold.

Analogous to $\gamma$, the orientations of the new cells' backbone axes are chosen randomly on division to decorrelate cell orientations.

Intercellular interactions are limited to steric repulsion between individual nodes, based on Hertzian contact theory\cs\cite{Hertz1882}.
Forces are computed from the node overlap $d$, as illustrated in \cref{pan:forceexternal}, with 
\begin{equation}
\label{F_cm}
    \mathbf{F}^{\text{ext}}\propto\Upsilon~ H(d)~ d^{3/2}~\mathbf{\hat{d}}
\end{equation}
where $\Upsilon$ is the Young's modulus of the cells, $H(\cdot)$ is the Heaviside function and $\mathbf{\hat{d}}$ is the normalized difference vector between the node centers. 
We choose $\Upsilon$ to be large enough to stay reasonably close to the incompressible limit (here: $10^5$, this choice is justified in \suppfigyoungsmodulus).

To keep forces at cell division continuous, we also use a Hertz-like force law with similar parameters for the internal spring, with
\begin{equation}
\label{eq:F_int}
    F\propto\Upsilon~\text{sgn}(\Delta b)~|\Delta b|^{3/2}
\end{equation}
where $\Delta b=b^{\mathrm{eq}}-||\mathbf{b}||$ is the current deviation of the backbone axis length from its rest length.
Node forces are combined into a force $\mathbf{F}^{\mathrm{cm}}$ acting on the cell's center of mass, a torque $\mathbf{T}$ acting on the backbone axis orientation, and a force $F^\mathrm{int}$ acting on the internal degree of freedom.

The evolution of the cells' degrees of freedom is described by overdamped equations of motion, where all forces are assumed to be instantaneously balanced by viscous effects. 
In this limit, velocities can be directly computed as the product of the relevant force with a mobility tensor, resulting in equations of motion such as $\dot{\vec{r}}^{\text{cm}}=\tens{\mu}^{\text{cm}}\vec{F}^{\text{cm}}$ for the center of mass position $\vec{r}^{\text{cm}}$ with mobility tensor $\tens{\mu}^{\text{cm}}$, and analogously for the orientation and the internal degree of freedom.
In our model, we use a numerical approximation for the mobilities of rods in a viscous fluid as derived by Tirado et al.\cs\cite{Tirado1984, Tirado1979, Tirado1980, Wensink2012}.

Simulations are initialized with eight cells, whose center of mass positions $\mathbf{r}^{\text{cm}}$ are set to the corners of a cube with a side length of 2 node diameters, with the initial growth progress $g$ and backbone orientation chosen randomly.
Due to their growth, cells will come into contact and begin interacting after a short time, quickly forming a spherical colony.
We integrate the system for $12$ generations (corresponding to 12 time units in the case $\alpha=1$).
Snapshots of the evolution of an initial condition are shown in \cref{pan:growthsnapshots}.
As there is no cell removal, we expect about $8\times2^{12}=32768$ cells to exist in the final state. However, this is a stochastic value since the growth rate and therefore also the times between divisions are stochastic.

All simulations are performed using \textit{InPartS}\cs\cite{InPartS}, an open-source framework for agent-based simulations of interacting particles developed in-house. 
Additional details on the model, its implementation and the simulations used for this work can be found in the \suppmat.

\section{Results}

\begin{figure}[ht]
    \centering
    \includegraphics[width=\figurewidth]{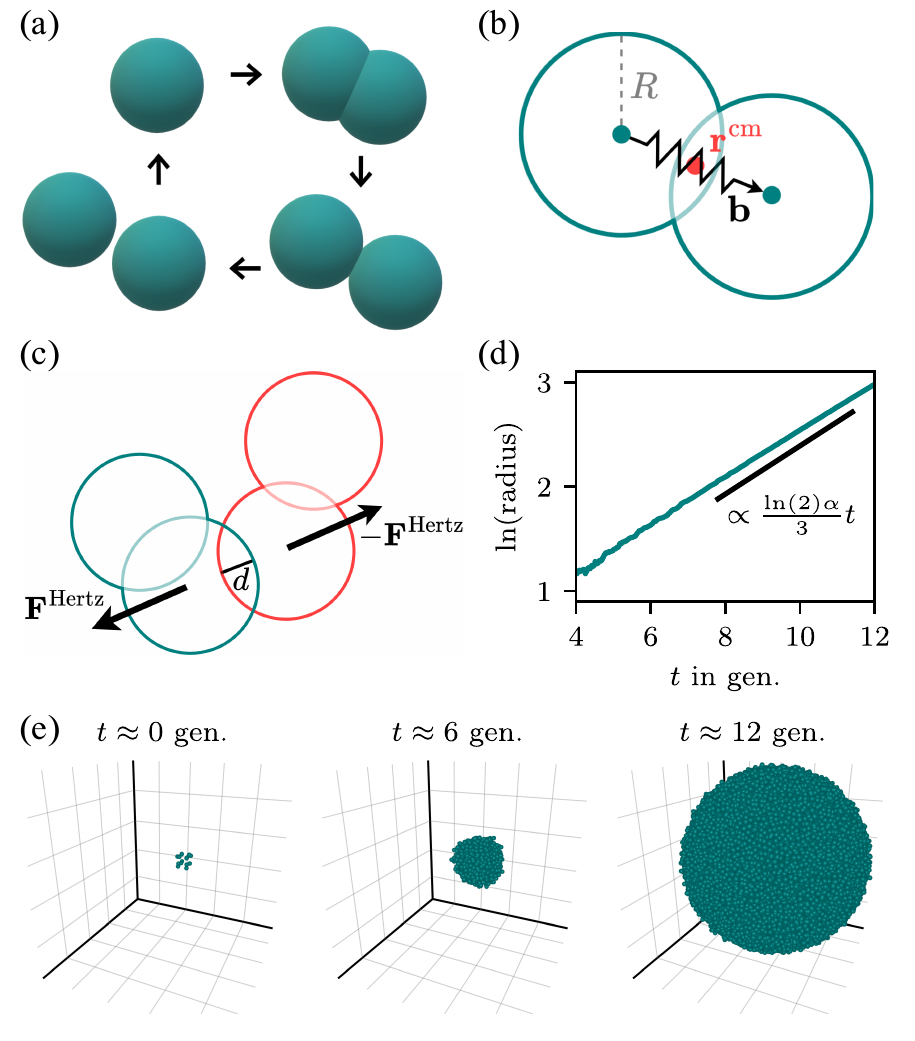}
    \mycaption{Cell model and macroscopic results. \panel{pan:division}~Life cycle of a cell: growth and division in the 3D dumbbell model in one generation.
    \panel{pan:forceinternal}~2D representation of the internal structure and internal forces of a cell. (see also \suppmat)
    \panel{pan:forceexternal}~2D representation of external forces by steric interaction repulsion between nodes of two cells.
    \panel{pan:radiustime}~Temporal evolution of the radius of the spheroidal cell colony. Black line indicates expected exponential scaling as described in the text.
    \panel{pan:growthsnapshots}~Snapshots of the spheroidal cell colony at different stages in time (see also \suppvidgrowth).}
    \label{fig:model_basicdynamics}
\end{figure}

We measure the radius of the spherical cell colony (\cref{pan:growthsnapshots} and \suppvidgrowth) by computing the particle number density as a function of distance from the center and setting a threshold based on a reference bulk density (see \suppmat for details).
\Cref{pan:radiustime} shows that this radius grows exponentially with time.
This growth is directly linked to the particle doubling rate $\alpha$ in the limit of a radially expanding incompressible material, where volume grows $\propto 2^{\alpha t}$ and therefore $R(t)\propto\exp(\alpha' t)$ with radial expansion rate $\alpha'= \frac{\ln(2)\alpha}{3}$.
As shown in \cref{pan:radiustime}, the numerical data is consistent with this scaling, confirming that we are working near the incompressible limit.

\begin{figure}[ht]
    \centering
    \includegraphics[width=\figurewidth]{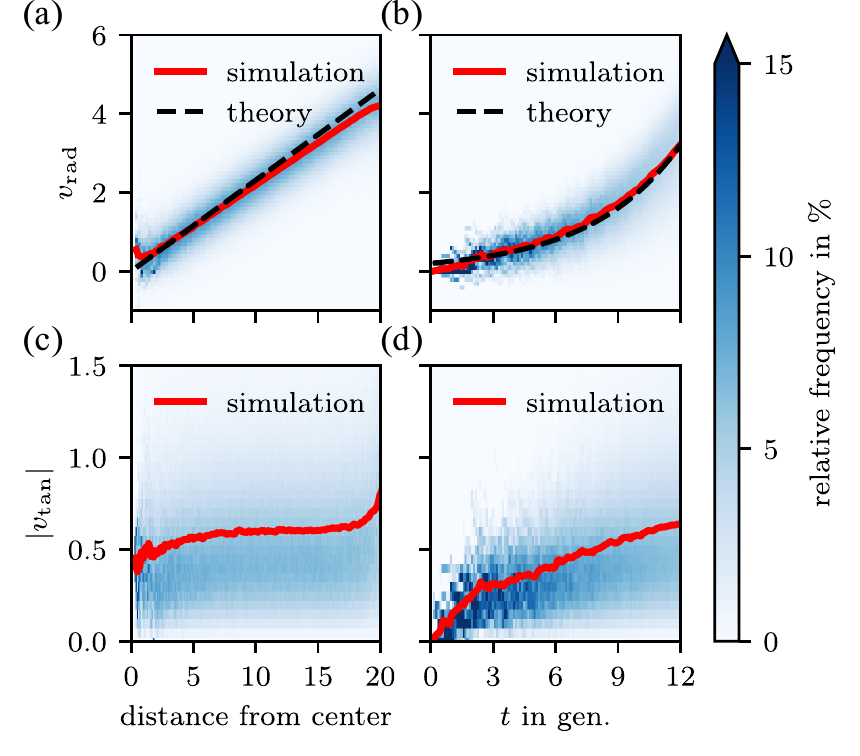}
    \mycaption{Characteristic signatures of cell velocities in different directions (see \suppmat and \suppfigvelocitydecomposition for technical details on the decomposition). \panel{pan:vrad_d}~Distribution, average and theoretical expectation \labelcref{eq:v_rad_r} of the radial velocity as a function of distance from the center.
    \panel{pan:vrad_t}~Distribution, average and theoretical expectation \labelcref{eq:v_rad_t} of the radial velocity against time. $R_0$ is determined from the last numerical data point.
    \panel{pan:vtan_d}~Distribution and average of the tangential velocity against distance.
    \panel{pan:vtan_t}~Distribution and average of the tangential velocity against time. In all panels, heatmaps represent histograms normalized for each column. The full spatio-temporal behavior of both velocity components is shown in \suppfigspeedheatmap.}
    \label{fig:velocities}
\end{figure}

As a first characterization of cellular motion, we measure the velocities of individual cells in the spheroid. In the radial direction, we expect motion to be dominated by the outward expansion flow. In contrast, from symmetry considerations, we expect tangential velocities to vanish on average. To account for this fundamental anisotropy, we decompose the velocities into radial and tangential components. Details of the velocity decomposition are discussed in the \suppmat.
Based on the same incompressibility assumption as above, we can easily calculate the global average of the radial velocity across all cells as a function of time, as well as the dependence of the radial velocity on radial position (independent of time):
\begin{align}
    \label{eq:v_rad_t}\langle v_{\text{rad}}\rangle_r(t)&=\frac{3}{4}\alpha' R(t)=\frac{3}{4}\alpha'R_0\exp(\alpha't)\\
    \label{eq:v_rad_r}v_{\text{rad}}(r)&=\alpha'r,
\end{align}
with $\alpha'$ as defined above.
The average radial velocities extracted from the simulation are consistent with these theoretical expectations, as seen in \cref{pan:vrad_d} and \cref{pan:vrad_t}. The imperfect incompressibility of the particles is visible as a slightly reduced slope in \cref{pan:vrad_d}. Likewise, the initial growth phase of sparse seed cells before forming a compact spheroid (cf. \cref{pan:growthsnapshots}) is visible as a deviation from exponential growth for $t\lesssim 3$ in \cref{pan:vrad_t}.

To characterize fluctuations around the vanishing average velocity in the tangential direction, we compute the absolute tangential speed $|v_\text{tan}|$.
We find it to have a wide distribution throughout the colony (blue heatmap in \cref{pan:vtan_d}), with an average which is fairly independent of the distance from the center of the spheroid (red line in \cref{pan:vtan_d}) and which increases slightly with time (\cref{pan:vtan_t}). The initial steep increase before $t\lesssim 3$ in \cref{pan:vtan_t} can again be associated with the very early compactification phase.
We also checked that there is no correlation between radial and tangential velocities of individual cells (\suppfigspeedcorrelation).
We conclude that tangential motion occurs everywhere in the colony, independent of position and radial velocities, as necessary rearrangements due to growth and division. This is also in line with $\langle|v_\text{tan}|\rangle$ being on the order of 1 cell node diameter per generation, which is much smaller than the radial velocity except very close to the center of the colony.

To quantify this non-trivial motion within the spheroidal cell colony, we calculate the mean squared displacement
\begin{equation}
     \text{MSD}(\Delta t)=\left\langle\left(\Delta \mathbf{x}_c(t_0,\Delta t)\right)^2\right\rangle_{t_0,c}
\end{equation}
by averaging the squared displacement $\left(\Delta \mathbf{x}_c(t_0,\Delta t)\right)^2$ after a lag time $\Delta t$ over all trajectories of all cells $c$ starting from all possible times $t_0$, only taking into account time intervals \emph{within} each cell's lifetime. This limits the MSD to (on average) one generation, but avoids choices regarding the treatment of splitting trajectories with shared history which can influence the statistics. While we are mostly interested in the characteristics of the non-trivial tangential motion, $\Delta \mathbf{x}_c$ for each cell can either be its tangential or total displacement.

Considering the total displacement, i.e., $\Delta\mathbf{x}_c(t_0,\Delta t)=\mathbf{r}_c^{\text{cm}}(t_0+\Delta t)-\mathbf{r}_c^{\text{cm}}(t_0)$, results in ballistic MSDs (\suppfigmsd), as one might expect from the dominance of radial advective motion described above. 
In contrast, taking into account only tangential displacements sampled from 2D trajectories of tangential motion (see \suppmat for details on their construction), we obtain the tangential MSD shown as a dark blue line in \cref{pan:msd} (lowest curve, $M=0$). It shows ballistic behavior only on short time scales, while the log-log slope $\kappa$ decreases towards $\kappa\gtrsim1$ for larger $\Delta t$, indicating long-term diffusive tangential motion (see \suppfigmsdslope for a plot of $\kappa$ vs. $\Delta t$). Even for $\Delta t$ towards $1\,\text{gen.}$, the absolute length scale $\sqrt{\text{MSD}}$ stays significantly below 1 cell node diameter, again consistent with motion caused by locally interacting, dividing cells.

\begin{figure}[ht!]
    \centering
    \includegraphics[width=\figurewidth]{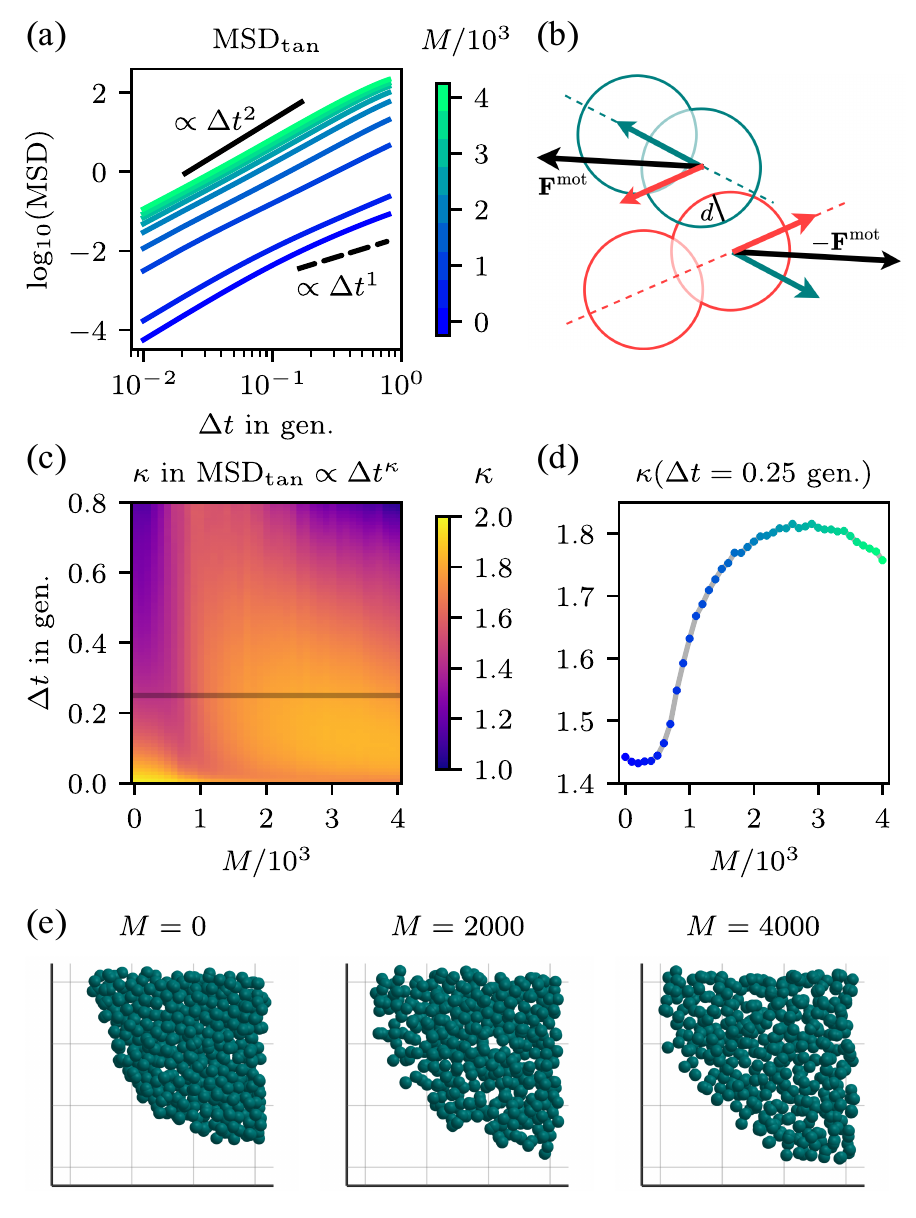}
    \mycaption{Cell motility enables tangential superdiffusion and loosening of the colony. \panel{pan:msd}~Double logarithmic representation of the tangential MSD as a function of the lag time $\Delta t$ for different motilities $M$. Insets depict slopes for comparison with ballistic and diffusive motion.
    \panel{pan:motilitymechanism}~Motility mechanism as an interaction between nodes of two cells. The dashed lines indicate backbone axis orientations.
    \panel{pan:msdslopes}~The slope $\kappa$ of the tangential MSD obtained from the double logarithmic representation for different lag times $\Delta t$ and different motilities $M$.
    \panel{pan:msdslopesend}~$\kappa$ at a lag time of $\Delta t=0.25~$gen. for different motilities indicates a transition at a motility $M\approx1000$.
    \panel{pan:motilitysnapshots}~A quarter of a slice through the center of the colony for different motilities at $t_{\text{max}}$ shows the decrease in cell number density for high motility.}
    \label{fig:msd_motility}
\end{figure}

So far, growth was the only provider of activity in our model, resulting in a strong radial expansion flow and cell-scale rearrangements to accommodate the growing and dividing matter. In this setting, all motion was caused by symmetrically expanding dumbbells of individual cells and volume exclusion interactions (cf. \cref{pan:division,pan:forceexternal}). While the two separating compartments of a cell could be viewed as self-propelled particles, their motion in the reference frame of the cell is by definition coupled to (and therefore cannot on average overcome) overall local volume expansion. In addition, the antialignment of this dipole-like pseudo-motility along the cell's backbone axis reduces the net effect on the cell's center of mass position. Consistent with our results so far, these effects limit the magnitude and nature of cell motion. From a theoretical perspective, it is therefore natural to consider a model extension with active forces causing true directed motion and to study how they modify the ballistic and diffusive regimes found so far. In addition, such motility is of high biological relevance in many growing systems from bacterial colonies to tumors\cs\cite{Wells2002, Wang2005,Hallatschek2023, Lopez2021, Gottheil2023}.

With cells being embedded in the dense three-dimensional aggregate, we assume that cell locomotion arises from cell-cell interactions which aim to move cells in the direction given by their backbone axes. In contrast to a fixed inherent self-propulsion velocity, this allows us to  maintain force balance in a physically consistent way. Following the scheme of steric interactions, these additional reciprocal interactions are implemented as forces between overlapping nodes of different cells $i$ and $j$:
\begin{equation}
    \mathbf{F}_i^{\mathrm{mot}}\propto M\cdot d \cdot H(d) \cdot (\hat{\mathbf{b}}_i-\hat{\mathbf{b}}_j),
    \label{eq:motility_force}
\end{equation}
With this, each node contributes a force aligned with the backbone axis direction $\hat{\mathbf{b}}$ of the cell it belongs to, and an equal and opposite force is applied to its interaction partner (see \cref{pan:motilitymechanism}).
Larger overlaps $d$ represent closer contact between cells and are therefore assumed to result in larger motility force. $M$ is the main control parameter modulating the overall strength of motility. Note that we do not add any explicit dynamics of the cell orientation (given by the backbone axis). Instead, reorientations continue to be caused purely by interactions with other cells.

In this extended model, we now re-evaluate the tangential MSDs for different values of the motility parameter $M$ (\cref{pan:msd}). As expected, we generally observe overall larger tangential MSDs for higher motilities (even the originally larger total MSDs increase significantly, see \suppfigmsd). A particularly noticable jump can be seen between $M=500$ and $M=1000$. In addition to the increase in magnitude, \cref{pan:msd} also shows a change in the behavior over time, with the initial ballistic regime persisting for longer $\Delta t$ compared to the non-motile case.
A finer scan of the motility parameter $M$ reveals a steep increase of the scaling exponent $\kappa$ of the MSD for a large range of different $\Delta t$ as soon as the motility parameter reaches, again, values around $M=1000$ (\cref{pan:msdslopes}). This sudden onset of strongly superdiffusive tangential motion at intermediate sub-generational time scales can be seen more quantitatively in \cref{pan:msdslopesend} for a lag time of $\Delta t=0.25~$gen., where $\kappa$ is virtually constant below $M\approx 500$ before exhibiting a rather sharp transition towards higher values. 
Even higher motilities lead to visible dispersion of cells, manifesting as a slightly increased colony size with lower cell number density (\cref{pan:motilitysnapshots}). This in itself will lead to less interaction (smaller overlaps $d$) and thus reduced motility forces for larger $M$, as seen in a slight decrease of $\kappa$ in \cref{pan:msdslopesend}. For now, we will focus on the regime of smaller $M$ where the transition occurs and this effect is weak, such that the motility force will be roughly proportional to $M$ itself, but further below, we will introduce a more suitable measure of motility force for larger values of $M$ which takes into account this feedback on cell arrangement. Note that, whenever we compare data between independent simulations (e.g., the snapshots in \cref{pan:motilitysnapshots}), we choose times relative to a well-defined time $t_{\text{max}}$ at which a certain cell number is reached towards the end of each simulation (see \suppmat for technical details). In this way, we always compare states corresponding roughly to the same number of cells, eliminating noise that is present simply due to small-number effects of randomly chosen cell cycle lengths in the initial stages of spheroid growth.

The apparent transition in the microscopic dynamics towards more directed motion at sub-generational time scales poses the question whether this goes along with changes in macroscopic and long-term dynamics.
A first qualitative change in the dynamics we detect is shown in \cref{pan:vradabs}: The radial speed abruptly loses the linear profile seen previously for the non-motile case (compare \cref{pan:vrad_d}), but only above a certain motility between $M\approx 700$ and $800$. In addition, a significant increase in the magnitude of radial speeds becomes visible only beyond this point.
As another quantitative measure, we examine the fraction of cells $f_{\text{in}}$ for which the radial velocity component $v_\text{rad}<0$, i.e., which move inwards. For simple radial expansion due to growth, we expect $v_r>0$ and thus $f_{\text{in}}\approx0$ everywhere except very close to the center of the spheroid. As \suppfiginwardsmovement shows, this is indeed the case below a certain motility threshold. However, $f_{\text{in}}$ quickly changes from $0$ to about $1/2$ in the range between $M=500$ and $M=1000$ depending on the radial position, indicating that the radial velocity becomes dominated and thus randomized by motility. 

The above analysis reveals a qualitative change of cell motion patterns in the spheroid at a specific motility threshold, pointing to a possible phase transition. However, it is not obvious how these global velocity changes translate into altered cell arrangements. This is further complicated by cell division, which makes the application of traditional measures like positional order parameters or neighbor tracking non-trivial. Since, for low $M$, the profiles in \cref{pan:vradabs} and \suppfiginwardsmovement still align with our naive expectation for a homogeneously expanding spheroid, one could suspect that cells are still confined in their local neighborhood in this weak-motility regime and motility-driven rearrangements can only occur for strong motility.

To characterize how cells move relative to each other over long times, we therefore extend our analysis from single cells to lineages, i.e., cells and their descendants.
More specifically, we detect all cells within a predefined spherical sample volume inside the colony at a certain sample time $t_0$ and then follow these cells and their descendants -- the ``sample lineages'' -- over some time interval $\Delta t$.
Taking into account that the sample volume itself grows together with the entire colony, we can distinguish between cases in which the sample lineages stay roughly within the growing sample volume and cases in which they escape and mixing throughout the colony occurs, as illustrated in \cref{pan:cagingmeasure}. In particular, we can characterize the dynamics entirely based on the dynamics of the proliferating cells themselves, without having to use auxiliary particles such as tracers, which can potentially deviate in their characteristics\cs\cite{MatozFernandez2017}.

We can formalize this idea by defining a mixing efficiency
\begin{equation}
    \eta_{M}(t_0,\Delta t)=2\frac{\langle||\mathbf{r}^\text{cm}_c(t_0+\Delta t)-\Bar{\mathbf{r}}(t_0+\Delta t)||
    \rangle_c}{R(t_0+\Delta t)}-\frac{1}{2},
    \label{eq:mixingefficiency}
\end{equation}
where $R$ is the colony radius as before, $\mathbf{r}^\text{cm}_c$ is the center position of cell $c$ and the average is taken over all cells in the sample lineages which exist at time $t_0+\Delta t$. $\Bar{\mathbf{r}} = \langle \mathbf{r}^\text{cm}_c\rangle_c$ is the overall average position of these (possibly dispersed) descendants and serves as a reference point to measure the spread of the sample lineages. We expect the relative spread $\langle||\mathbf{r}^\text{cm}_c-\Bar{\mathbf{r}}|| \rangle_c/R$ to stay constant if a compact sphere made up of the sample lineages simply expands together with the entire spheroid, whereas it will increase if mixing occurs. For simplicity, $\eta_{M}$ is normalized here for a specific sample volume radius of $R/3$: The edge cases are a uniform distribution of cells within the sample volume (no mixing), resulting in a relative spread of $\frac{1}{4}$, and a uniform distribution of cells within the entire spheroid volume (perfect mixing), resulting in a relative spread of $\frac{3}{4}$. After normalization of the relative spread according to these limiting values, we obtain the mixing measure in \cref{eq:mixingefficiency}, which starts out at $\eta_{M}(t_0,0)\approx0$ at sample volume initialization, and where, in general, the edge cases now correspond to values of $\eta_{M}=0$ and $\eta_{M}=1$, respectively.

\cref{pan:mixingefficiency} shows the mixing efficiency as a function of motility after different lag times $\Delta t$, all starting from $t_{\text{max}}-3~$gen., respectively. These curves confirm the limiting cases of perfect confinement for small motilities and perfect global mixing for large enough motilities.
Most importantly, however, mixing only seems to set in at a certain value of the motility parameter.
Between the three values of $\Delta t$, only the steepness of the transition varies, whereas the onset remains fairly unchanged.
This behavior can also be confirmed visually in \suppvidsmixing, where no mixing of the sample lineages with the rest of the spheroid is observed both for $M=0$ and $M=500$, but $M=1000$ and up show significant mixing. It should be emphasized that we have a homogeneous system in which all cells, both the marked cells of the spreading sample volume and those in the rest of the spheroid, have identical properties. Thus, the mixing measure as well as its visual representation indicate a global phenomenon.

\begin{figure}[ht]
    \centering
    \includegraphics[width=\figurewidth]{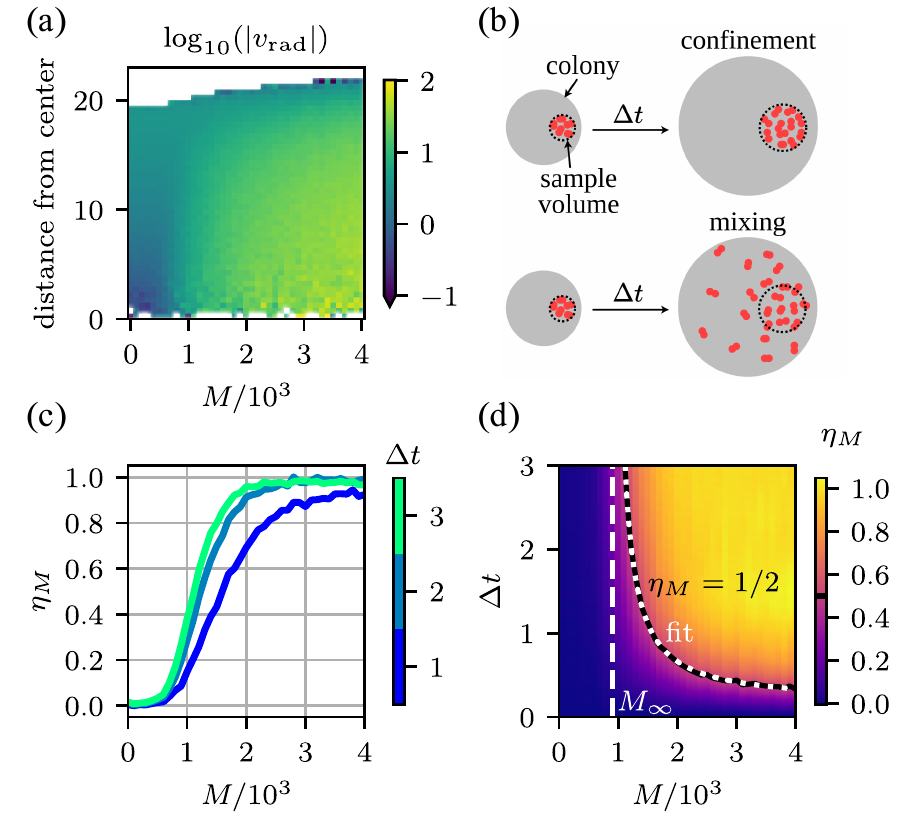}
    \mycaption{Motility-induced mixing. \panel{pan:vradabs}~Spatial profile of the radial speed after $t_{\text{max}}$ for different motilities $M$. One can observe that for larger motilities, the radial speed is almost two orders of magnitude higher than for the non-motile case. Also it is larger in the inside of the colony. This goes along with cells moving both inwards and outwards. 
    \panel{pan:cagingmeasure}~Tracking cells from a sample volume and their descendants over time with respect to a growing reference volume serves as a measure to differentiate between confinement and mixing.
    \panel{pan:mixingefficiency}~Mixing efficiency at different motilities for three different lag times $\Delta t$ after a starting time $t_{\text{max}}-3~$gen. Data is averaged from 6 non-overlapping sample volumes (see \suppmat). See \suppvidsmixing for the spatiotemporal dynamics of the sample lineages at selected motilities.
    \panel{pan:uncagingtime}~Color-coded representation of the mixing efficiency as a function of motility and lag time. The black line is a contour at a mixing efficiency $\eta_M=1/2$.
    The white dotted line is a shifted hyperbola fitted to the contour line. It exhibits asymptotic behavior at $M_{\infty}$, indicated by the white dashed line.
    } 
    \label{fig:uncaging}
\end{figure}%

This evidence suggests that waiting for longer lag time $\Delta t$ cannot compensate for low motility below a certain threshold, raising the question whether the existence of this regime be substantiated quantitatively. To assess the characteristic time scale of mixing more directly, we therefore calculate the mixing efficiency $\eta_M$ for arbitrary lag times $\Delta t\in[0~\text{gen.},3~\text{gen.}]$. The heatmap in \cref{pan:uncagingtime} shows $\eta_M$ as a function of the motility and lag time. A contour line at an arbitrary value of $\eta_M=1/2$ (black line) indicates the time scale of mixing as a function of motility. This time scale indeed appears to exhibit asymptotic divergence at a finite non-zero value $M_{\infty}$ of the motility parameter. Such behavior would imply the existence of a motility threshold $M_{\infty}$ below which mixing never reaches $\eta_M=1/2$, irrespective of the lag time. To determine whether the scaling behavior of $\eta_M$ is compatible with this, we fit a shifted hyperbola of the form
\begin{equation}
    \label{eq:shifted_hyperbola}\Delta t|_{\eta_M=1/2}(M)=A\cdot (M-M_{\infty})^{\beta}+C
\end{equation}
to the contour line, shown as a white dotted line in \cref{pan:uncagingtime} (see \suppmat for details). The double-logarithmic plot in \suppfigmixingheatmaploglog further confirms the appropriateness of this functional form, which indicates an asymptote at $M_{\infty}\approx900$.
This is qualitatively reminiscent of a glass transition: A diverging time scale of mixing at a finite, non-zero motility $M_{\infty}$.

These observations raise the question how the mixing dynamics and in particular the suppression of mixing can be understood both conceptually and quantitatively. At what speed do individual cells actually move through their local environment at a given value of $M$, beyond proliferation-induced rearrangements? What role does the expanding nature of the system play in the ability of cells to spread over large distances?
To answer these questions, we build a phenomenological model of the mixing dynamics that still allows us to extract relevant parameters that can be compared with direct measurements.
In this effective description, we approximate the spreading of cells from the sample volume with a continuum model where a density, originally concentrated in the sample volume, undergoes a diffusion process. The two parameters are thus the radius $Q$ of the initial uniform sphere and a diffusivity $D$. We first consider regular, non-expanding space. In this case, the probability density function $P(r,t)$ can be found simply by a convolution of the initial condition with the heat kernel\cs\cite{Carslaw1959, Lovering1935, Ingersoll1913} (see \suppmat).
From this distribution, the time course of the absolute spread can be computed as
\begin{align}
    &\langle r\rangle(t)=\iiint_VrP(r,t)~\dd V=\\
    &\frac{3}{8Q^3}\left[\frac{\sqrt{u}Q(u+2Q^2)}{\sqrt{\pi}\exp\left(\frac{Q^2}{u}\right)}-\frac{u^2-4uQ^2-4Q^4}{2}\text{erf}\left(\frac{Q}{\sqrt{u}}\right)\right],
    \label{eq:model_spread}
\end{align}
where $u=4Dt$. Exponential growth of space can then be accounted for by rescaling time \mbox{$t\rightarrow\frac{1-\exp(-2\alpha't)}{2\alpha'}$}, and consequently $u$ (see \suppmat for derivation). The result is a system in which the effective diffusivity decreases over time (while the domain does not grow), equivalent to a constant diffusivity in an expanding space where length scales have been rescaled to be kept constant. In particular, \cref{eq:model_spread} with this rescaled $u$ allows us to calculate the relative spread for a central initial sample volume with $Q=R_*/3$ using \cref{eq:mixingefficiency}, where we can substitute $R(t)=R_*=\text{const.}$ due to the nature of the rescaling and $R_*$ is the colony radius at sample volume initialization. We fit the resulting expression for $\eta_M$ to data from simulations of the different motility values, where the only fit parameter is the diffusivity $D$ (\cref{pan:model_eta}). This provides a relation between the motility parameter $M$ and the effective diffusivity $D$ which is shown in \cref{pan:model_D}.
Additionally, the transform itself contains useful information on the limiting behavior of this process: Taking $t\rightarrow\infty$ in the transformed $u=\frac{2D}{\alpha'}(1-\exp(-2\alpha't))$, we immediately see that $u$ can reach a maximum value of $2D/\alpha'$, which also limits the maximum spread according to \cref{eq:model_spread} and thus the maximum mixing efficiency that can be reached for a given diffusivity $D$. Let $u_{1/2}$ be the value of $u$ which yields a mixing efficiency of $\eta_M=1/2$, we can thus define
\begin{equation}
    D_{\infty}=\frac{u_{1/2}\alpha'}{2}
\end{equation}
in analogy to $M_\infty$, i.e., the critical diffusivity below which, geometrically, no mixing beyond $\eta_M=1/2$ is possible. The necessary value of $u_{1/2}$ can be found numerically from \cref{eq:model_spread} and \cref{eq:mixingefficiency}, and only depends on $Q$ and thereby the radius $R_*$ of the colony at sample volume initialization. The resulting limiting diffusivity $D_\infty$ shown in \cref{pan:model_D} indeed indicates that mixing should not be able to surpass $\eta_M=1/2$ below $M\approx1000$, close to the value we found before.

Having described the dynamics in terms of an effective diffusivity $D$, the question remains how this diffusivity is related to microscopic locomotion of cells through the spheroid. While our control parameter $M$ modulates motility, the driving forces of motility also depend on the emerging cell configuration, such as the number and strength of cell-cell contacts, rendering the relationship of $M$ and $D$ (\cref{pan:model_D}) nontrivial. A more intuitive description for the source of diffusivity would be in terms of an effective self-propulsion velocity. Conceptually consistent with the previously described ballistic-to-diffusive MSDs, a description that captures the essential features of our motility mechanism is that of a free athermal active Brownian particle (ABP) with a constant active self-propulsion velocity $v$ and rotational diffusivity $D_r=1/\tau_r$ (caused by changes in cell-cell contacts). To take into account the randomization of orientation at cell division, we further assume tumbles after run times of fixed duration $T$ corresponding to the cell cycle length. We can derive a long-term diffusion coefficient $D$ associated with this type of motion by building on general results for intermittently self-propelled particles derived by Datta et al.\cs\cite{Datta2024}, resulting in a relation known from ABPs:
\begin{equation}
    D=\frac{1}{3}v^2\tau_{\text{eff}}
    \label{eq:abp_and_tumble}
\end{equation}
with an effective persistence time (see \suppmat)
 \begin{equation}
       \tau_{\text{eff}}=\tau_r\left(1+\frac{\tau_r}{T}\left(\exp\left(-\frac{T}{\tau_r}\right)-1\right)\right).
       \label{eq:tau_eff}
\end{equation}
The intra-generational $\tau_r$ for each motility can be obtained from the initial decline of the orientation autocorrelation function (OACF, see details in \suppfigaacf), for which the relevant orientation is given by the current direction of motion (consistent with the free ABP analogy). Using a cell cycle length of $T=1\,\text{gen.}$, the effective persistence time $\tau_{\text{eff}}$ resulting from \cref{eq:tau_eff} is shown for each value of the motility parameter in \cref{pan:model_tau}\cs\footnote{Note that the true cell cycle length (run time between tumbles) does not have a fixed value $T$ but is distributed according to the growth rate distribution in our simulations. However, considering the true run time distribution (\suppmat, particularly \suppfigtumblingabp) only leads to negligible differences, so we consider the simpler case of a $\delta$ distribution here.}. Given these values of $\tau_{\text{eff}}$ and those of the diffusion coefficient $D$ found in \cref{pan:model_D}, we can use 
\cref{eq:abp_and_tumble} to calculate the effective self-propulsion velocity $v=\sqrt{3D/\tau_{\text{eff}}}$ for each value of $M$. Assuming this velocity is isotropically distributed on the sphere, the corresponding tangential component will be $\pi v/4$, which we can compare with the actually observed tangential motion in the spheroid simulations (\cref{pan:model_v}). The excellent match between the self-propulsion velocity inferred from the ABP model and the true tangential velocities, particularly in the transition region, substantiates the validity of our phenomenological description and thus the conclusions drawn from it.
Note that both velocities examined so far as well as the diffusivity (\cref{pan:model_D}) characterize the motion actually performed by cells. Their relatively late onset with respect to $M$ is remarkable, since abundant cell-cell contacts in this dense system should generate motility forces according to \cref{eq:motility_force} for any finite value of $M$. To confirm this, we can compute the tangential velocity that would arise from motility forces only, in the absence of steric repulsion, showing its immediate onset for any $M>0$ (\cref{pan:model_v}). We call this potential velocity the average active velocity $\Bar{v}_a$, not to be confused with the effective self-propulsion velocity examined above. Steric interactions in the densely packed spheroid therefore appear to suppress locomotion to a certain extent, resulting in the absence or lower magnitude of observed tangential velocities (or, equivalently, inferred active self-propulsion velocities). Note that $\Bar{v}_a$ can be seen as a parameter of the system, similar to $M$, only that it already accounts for other effects that modify the motility, \cref{eq:motility_force}. Using this effective parameter (which can also be more closely related to active velocities ordinarily considered in active matter) instead of $M$ will allow us to compare scenarios where cell-cell contacts are modified due to changes in parameters unrelated to motility. 

\begin{figure}[ht!]
    \centering
    \includegraphics[width=\figurewidth]{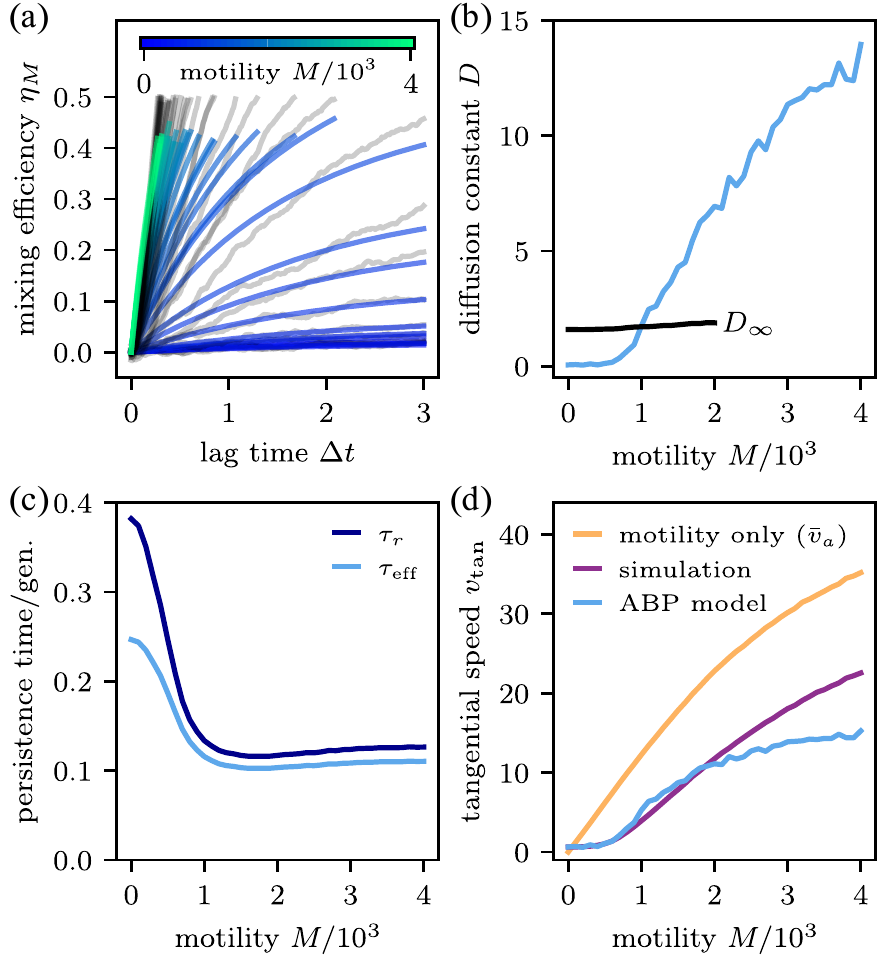}
    \mycaption{Phenomenological model of a diffusing uniform sphere of non-interacting tumbling athermal ABPs in an exponentially growing domain.
    \panel{pan:model_eta}~Time series of the mixing efficiency $\eta_M$ for different motility values (transparent gray lines) with the respective fits of the mixing efficiency derived from a model of the diffusion of a uniform sphere on an exponentially growing domain.
    \panel{pan:model_D}~Fit parameter $D$ as a function of the motility. Black line indicates critical diffusivity $D_{\infty}$ at which the expansion suppresses diffusion such that no mixing to $\eta_M=1/2$ is possible. This is reached around a motility of $M_{\infty}\approx1000$.
    \panel{pan:model_tau}~Persistence time of rotational dynamics. Dark blue line is based on fits of exponential decays on the orientation autocorrelation, obtained from the time series of velocity directions between division events (\suppfigaacf). Light blue line combines this with additional tumbles according to \cref{eq:tau_eff}.
    \panel{pan:model_v}~Tangential velocities from motility forces only, in the absence of steric interactions (yellow line, active velocity $\bar{v}_a$) increase with the motility parameter, while tangential velocities from the simulation (purple line) begin to increase only above a certain motility threshold. The inferred self-propulsion velocity in the phenomenological model (light blue line) largely captures this.}
    \label{fig:abpmodel}
\end{figure}

The failure of a potential active velocity to actually manifest in the dynamics does not, in itself, reveal the physical origin of this effect. However, the other non-equilibrium activity, proliferation, is a plausible candidate due to the positive pressure it induces (i.e., on-average inward forces on cells), which might prevent lineages from overcoming their local confinement. This would mean that varying the growth rate should alter the strength of these confining forces and thereby the critical motility force to overcome them.
To test this hypothesis, we performed simulations with different bulk doubling rates $\alpha$ between $0.1$ and $1.0$. According to the line of thought above, with smaller growth rates inducing a lower pressure, we expect that smaller motility forces are sufficient to actually observe motion. Since a reduced growth rate also leads to less contact between cells, it will also change the average motility force for the same value of $M$, such that comparisons of the mixing transition in terms of the parameter $M$ (as seen in \cref{pan:uncagingtime}) become meaningless. Therefore, we first measure the average active velocity $\Bar{v}_a$ (which corresponds to a certain effective motility force) as introduced above to enable a fair comparison between different scenarios.
Plotting $\Bar{v}_a$ as a function of motility parameter $M$ for the different bulk doubling rates $\alpha$ (\cref{pan:active_velocity}),
we can then use these calibration curves\cs\footnote{in reality, we use low-degree polynomial fits which capture the data extremely well but eliminate any noise that might lead to nonmonotonicity, see \suppfigactivevelocityfit} to convert our abstract motility parameter $M$ into average active velocities for different growth rates, and examine the time necessary to reach $\eta_M=1/2$ as a function of this more meaningful parameter for all growth rates (\cref{pan:contourlines}). This shift between different curves already shows that the active velocities necessary for to reach $\eta_M=1/2$ within the same amount of time decreases with the growth rate, suggesting that it is easier for cells to move through a more slowly growing spheroid.

 We can then fit shifted hyperbolas to these contours again as in \cref{eq:shifted_hyperbola}, only now as a function of $\bar v_a$, to determine the critical active velocity $\bar v_{a,\infty}$ at which the time scale $\Delta t|_{\eta_M=1/2}$ for reaching $\eta_M=1/2$ diverges (see \suppfigbetafitparam for details), as indicated by dots at the top of \cref{pan:contourlines}.
In a similar way, we can examine the initial onset of actually observed speeds as in \cref{pan:model_v}, but again as a function of $\bar v_a$ instead of $M$, and thereby determine the minimum active velocity $\bar v_{a,\text{onset}}$ necessary to induce appreciable tangential motion for different growth rates (see \suppfigvelocityonset). Both critical active velocities are plotted in \cref{pan:critical_velocities}.

\begin{figure*}[ht]
    \centering
    \includegraphics[width=0.8\textwidth]{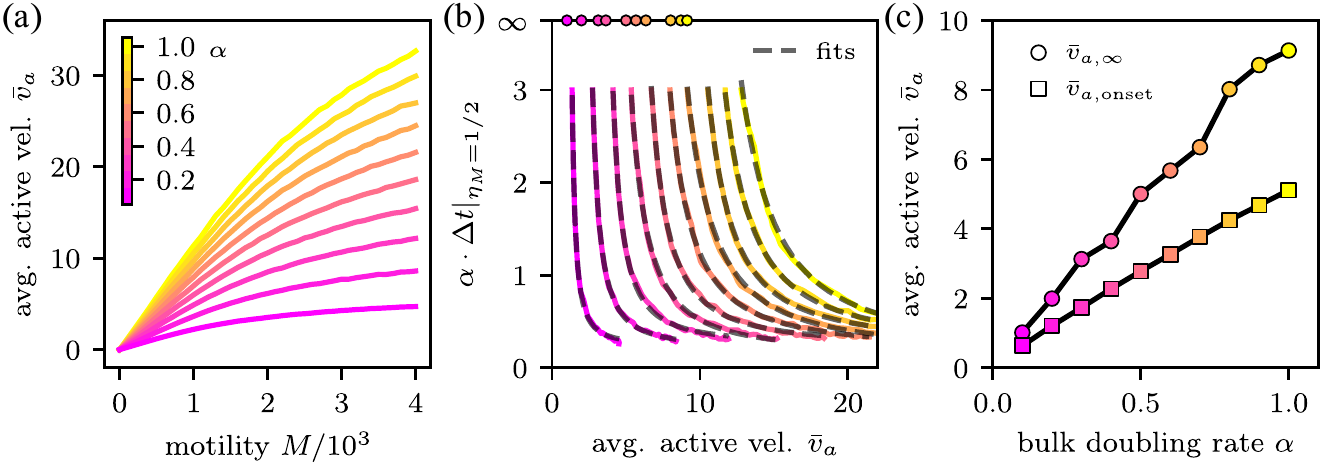}
    \mycaption{Motility-induced mixing transition for different growth rates. \panel{pan:active_velocity}~Profiles of the average active velocity analogous to the corresponding yellow line in \cref{pan:model_v} for different bulk doubling rates $\alpha$.
    \panel{pan:contourlines}~Contour lines representing the lag time (here scaled with $\alpha$, such that one time unit corresponds to one true generation at the respective doubling rate) necessary to reach mixing efficiency $\eta_M=1/2$, analogous to \cref{pan:uncagingtime}, but in terms of $\bar v_a$, for different bulk doubling rates $\alpha$. Dots at the top indicate the critical active velocity $\bar v_{a,\infty}$ at which the mixing time diverges, obtained from shifted hyperbola fits.
    \panel{pan:critical_velocities}~Critical active velocity $\bar v_{a,\infty}$ from panel (b) as a function of the bulk doubling rate $\alpha$ (circles), compared to the minimum active velocity $\bar v_{a,\text{onset}}$ at which the tangential speed $v_{\text{tan}}$ begins to increase with motility (squares; see \suppfigvelocityonset for details).}
    \label{fig:growthrate}
\end{figure*}

Already from \cref{fig:abpmodel} we learned that the mixing transition is in fact governed not by one but by two effects: First, active velocities below a certain threshold $\bar v_{a,\text{onset}}$ are \emph{suppressed} and manifest neither in increased actually observed velocities (\cref{pan:model_v}) nor an effective diffusion (\cref{pan:model_D}) on a lineage level. Second, even for $\bar v_a > \bar v_{a,\text{onset}}$, \emph{expansion} prevents active locomotion from reaching arbitrarily large values of $\eta_M$, indicated by the existence of a minimum diffusivity $D_{\infty}$ necessary to reach $\eta_M=1/2$, which corresponds to a minimum active velocity $\bar v_{a,\infty}$ as determined in \cref{pan:contourlines}. Both effects act in concert to create the motility-induced mixing transition. While velocity suppression below $\bar v_{a,\text{onset}}$ causes dynamical arrest on the lineage level at a finite active velocity, the expansion effect is responsible for the diverging mixing time scale when motility already exceeds this threshold. Therefore, velocity suppression shifts the necessary active velocity to larger values. In \cref{pan:critical_velocities} we can see that, e.g., for the default bulk doubling rate $\alpha=1.0$, an active velocity of $\bar{v}_a\approx5$ is required before the observed onset of $v_{\text{tan}}$ and an additional active velocity of $\bar{v}_a\approx4$ to reach a diffusivity sufficient to enable mixing up to $\eta_M=1/2$.

The fact that $\bar v_{a,\text{onset}}$ increases with the growth rate supports our hypothesis that cells need to overcome the confinement created by their environment due to growth-induced pressure. Interestingly, both critical active velocities increase roughly linearly with the growth rate, suggesting that both contributions remain essential across a wide range of growth rates instead of dominating each other in some limit. However, it should be noted that their relative contributions can vary with the extent of mixing, i.e., the specific value of $\eta_M$ that one considers successful mixing. This is because the critical diffusivity $D_\infty$ (and thereby the actual velocity) required to overcome the expansion effect depends on the target value of $\eta_M$ via $u_{1/2}$, with a lower required velocity for lower $\eta_M$. However, any actual velocity only manifests beyond $\bar v_{a,\text{onset}}$. Therefore, the same active velocity is required to overcome velocity suppression, while the additional active velocity required to overcome the expansion effect varies with the extent of mixing.

\section{Discussion}

Our theoretical study set out to characterize the collective dynamics in a growing spheroidal cell colony. It can be seen in the context of a number of recent studies in the field of active matter which have provided novel perspectives on cell motion in biological systems by relating cellular activity -- proliferation or motility -- to phenomena such as fluidization, glass or rigidity transitions\cs\cite{Sinha2020, Bi2016, Huang2022, Malmi18, Ranft2010, Hannezo2022}.
Here, we built an agent-based computational model that allowed us to study the dynamics and self-organization
arising from minimal ingredients, namely growth, division, motility and steric repulsion. We deliberately refrained from adding explicit growth regulation, resulting in simple scaling laws for volume expansion and radial velocity in the non-motile case. It is therefore consistent with the early stages of a biological colony when chemical gradients due to diffusion limitation or gene-regulatory-induced heterogeneities have not yet developed, but could also be applicable to future artificial systems in which such regulation is absent. However, more importantly, our model can serve as a baseline to delineate the effects of additional mechanisms in theory and experiments.

Even without motility, we observed tangential motion of cells in the spheroid.
This is a purely emergent dynamical feature and may be understood as cell-scale rearrangements due to local growth and division, as evidenced by its small, position-independent magnitude and approximately diffusive nature in contrast to ballistic radial motion.
Although we consider the cell center for our MSD calculation, which only moves due to the interaction with the bath of growing cells and not due to growth of the cell itself, tangential MSDs show ballistic scaling on very short sub-generational time scales, consistent with the deterministic single-cell growth process in this athermal system.
Towards a time scale of one generation, the system randomizes sufficiently to yield approximately diffusive tangential motion. 
Our mixing analysis showed that lineages stay perfectly confined in their local neighborhoods without motility, which seems to contradict the locally diffusive motion, which indicates a fluidized system. However, this can be resolved by taking into account two aspects: First, the two characterizations apply to different entities, i.e., we have to differentiate between single-cell dynamics and lineage dynamics, which also means that analogies to glassy dynamics must necessarily remain on a conceptual level. Second, proliferation of cells creates additional space in which this motion can take place without leaving the (expanding) neighborhood. The coexistence of both features can therefore be considered a distinct feature of systems with proliferation and isotropic volume expansion. We also think that our results do not contradict the superdiffusive motion and fluidization seen in a tumor model with growth regulation\cs\cite{Malmi18}, since our tangential MSDs exclude the more directed radial motion and the mixing analysis is agnostic to cell rearrangements within the sample lineages. 
Interestingly, tangential diffusion as passive translocation on the surface of tumors has been considered in mathematical reaction-diffusion models before\cs\cite{Tracqui_1995}.

On the sub-generational time scale we revealed the emergence of a transient superdiffusive regime of tangential motion induced by motility. The sudden change in the scaling behavior of the tangential MSD at a finite motility is accompanied by equally abrupt changes in the instantaneous radial velocities. The additional directed motion on short time scales can be interpreted as a manifestation of motility which is suppressed below the transition point due to the dense packing of cells. We were able to substantiate this interpretation by comparing the potential motility due to an average active velocity with the actually observed velocity, whose onset occurs around the same motility of $M\approx500$ (\cref{pan:model_v}) as the transition point in the tangential MSDs (\cref{pan:msdslopesend}). Beyond the transition, cells are then able to use this directed motion to escape from their local neighborhoods. We could further underscore the crucial role of steric interactions by varying the growth rate and thereby modulating their strength, which led to changes in the critical active velocity in the expected direction. An interesting aspect of the velocity suppression is its reminiscence of contact inhibition of locomotion, which is an active biological regulatory process\cs\cite{Stramer2016}, whereas similar phenomenology here is caused by a purely mechanical mechanism.

The suppression of motion at a non-zero motility is an emergent phenomenon that is similar to a glass transition and has previously been observed in biological systems\cs\cite{Malmi18, Bi2016, Boocock2023, Angelini2011}, also explicitly in growing dense active matter\cs\cite{Malmi18, Tjhung2020}. While Tjhung et al. found glass-like dynamics, manifested by different phenomena, such as aging, the regulation of growth by a total energy density played a critical role in their system\cs\cite{Tjhung2020}. This can be contrasted with our model, which does not consider growth rate control mechanisms and all local dynamics take place on a background of constant expansion. More specifically, an onset of motion beyond a critical motility has already been related to a glass transition in self-propelled voronoi models\cs\cite{Bi2016}, and similarly, fluidization at a critical magnitude of a random traction force has been reported in a random traction vertex model\cs\cite{Amiri2023}. However, our baseline case differs in the sense that cell rearrangements due to proliferation fluidize the system even before the transition.

The most striking feature of our system is the existence of a mixing transition characterized by a diverging timescale at finite motility. It is important to note that directed motion in the spheroid, once it happens, is still randomized on long time scales and does not result in collective motion or even rotation which has been observed in other systems\cs\cite{Ascione_2023}. This allowed us to quantitatively capture the mixing dynamics using a phenomenological model based on diffusion in an exponentially expanding sphere, where diffusion is imagined to result from the motion of a tumbling athermal active Brownian particle. It revealed that the exponentially expanding nature of the system, causing all points to continuously distance from each other reminiscent of Hubble expansion\cs\cite{DellArciprete2018, Hubble1929}, has a significant contribution to the existence of a diverging mixing time scale and leads to a critical diffusivity necessary to overcome expansion (\cref{pan:model_D}). The inferred diffusivity present at different motilities in the spheroid indeed reaches this critical value in the motility range that also exhibits the mixing transition (compare \cref{pan:uncagingtime} and \cref{pan:model_D}). Both this and the quantitative match between the inferred self-propulsion velocities and measured ones (\cref{pan:model_v}) validate the description, which integrates all aspects of the dynamics---in particular the above-mentioned velocity suppression and the expansive growth of the spheroid---and can explain their mutual interactions.

Here, we did not consider apoptosis, i.e., removal of cells, which has been shown to cause fluidization on its own\cs\cite{Ranft2010}, dependent on division and death mechanisms\cs\cite{Reddy2022}. In systems with apoptosis and additional adhesion, tissue fluidization also occurs independently from the existence of a glassy phase\cs\cite{MatozFernandez2017}. It would be therefore interesting to systematically study how the specific, multi-stage, route to mixing found here is modulated by other processes such as apoptosis or cell adhesion, the latter of which has also been found to be important for cancer invasion\cs\cite{Labernadie2017}.%

Besides the analogy with rigidity and glasses, the mixing dynamics may also have practical consequences, as it enables transport within the colony, including between the bulk and its surface. This could be relevant for the escape of specific cells in the case of phenotypic or genetic heterogeneity. For example, in cancer research, such intratumor heterogeneity is a diligent subject\cs\cite{Waclaw2015, Ahmed2017}. It might therefore be worthwhile to study how important collective effects are for the behavior found here, i.e., whether and how it translates to individual motile sub-populations or, in general, heterogeneous motility or growth.

Experimental investigations based on our findings could try to identify both the onset of (increased) tangential superdiffusive motion, which itself has already been found in similar systems\cs\cite{Raghuraman2022}, and the critical tangential velocity of cells required for mixing in growing multicellular spheroids. 
While calculating numbers characterizing the strength of activity, such as the P\'eclet number, is difficult in our system (due to the different levels of description and the presence of multiple activities), it is interesting to note that both critical velocities we detect are roughly proportional to the growth rate (\cref{pan:critical_velocities}). Non-dimensionalizing the velocities with the growth rate therefore yields approximately constant velocities in units of cell diameters per generation. One additional avenue for future theoretical research could therefore be to determine which properties of the system determine this quasi-universal value.
Additionally, this value on the order of $10~$cell diameters per generation is comparable to typical velocities observed in experiments (see \suppmat for estimation), e.g., $(14\pm8)~$cell diameters per generation before and $(30\pm15)~$cell diameters per generation after the epithelial-mesenchymal transition (EMT)\cs\cite{Quinsgaard2024}, which is a known driver of cell motility. Therefore, the regime explored in our study seems relevant for real-world scenarios.

\begin{acknowledgements}
We thank Jonas Isensee, Ramin Golestanian, Timo Betz, Matthias Kr\"uger, Leif Peters, Yoav Pollack and Peter Sollich for valuable discussions and input.
\end{acknowledgements}

\section*{Data availability}
Example code for simulations and data analysis methods used in this publication are available for download \href{http://hdl.handle.net/21.11101/0000-0007-FE41-2}{here}. It is based on our in-house simulation framework \textit{InPartS}\cs\cite{InPartS} and the \href{http://hdl.handle.net/21.11101/0000-0007-FE13-6}{model repository} which contains the specific cell model used here.

\interlinepenalty=10000

\end{document}